\documentclass[conference]{IEEEtran}
\IEEEoverridecommandlockouts
\usepackage{cite}

\usepackage{amsmath,amssymb,amsfonts}
\usepackage{algorithmic}
\usepackage{graphicx}
\usepackage{textcomp}
\usepackage{xcolor}
\usepackage[ruled,linesnumbered]{algorithm2e} 
\usepackage[nolist]{acronym}
\usepackage{soul}
\usepackage{marginnote}
\usepackage{url}
\usepackage[inline]{enumitem}
\usepackage{tabularx}
\usepackage{amsthm}
\usepackage{tikz}
\usepackage{tikz}
\usepackage{hyperref}
\hypersetup{
    colorlinks=true,
    linkcolor=blue,
    filecolor=magenta,      
    urlcolor=cyan,
    pdftitle={Overleaf Example},
    pdfpagemode=FullScreen,
    }
\usepackage{cleveref}

\newcommand*{\thead}[1]{\multicolumn{1}{c}{\bfseries #1}}

\newcommand{\optional}[1]{%
  {\color{gray}#1%
  }%
}
\renewcommand{\optional}[1]{}

\begin{document}

\title{
  Energy-Efficient Deployment of Stateful FaaS Vertical Applications on Edge Data Networks
}

\author{
\IEEEauthorblockN{Claudio Cicconetti}
\IEEEauthorblockA{\textit{IIT, CNR} --
Pisa, Italy \\
c.cicconetti@iit.cnr.it}
\and
\IEEEauthorblockN{Raffaele Bruno}
\IEEEauthorblockA{\textit{IIT, CNR} --
Pisa, Italy \\
r.bruno@iit.cnr.it}
\and
\IEEEauthorblockN{Andrea Passarella}
\IEEEauthorblockA{\textit{IIT, CNR} --
Pisa, Italy \\
a.passarella@iit.cnr.it}
}


\IEEEtitleabstractindextext{%
\begin{abstract}
5G and beyond support the deployment of vertical applications, which is particularly appealing in combination with network slicing and edge computing to create a logically isolated environment for executing customer services.
Even if serverless computing has gained significant interest as a cloud-native technology its adoption at the edge is lagging, especially because of the need to support stateful tasks, which are commonplace in, e.g., cognitive services, but not fully amenable to being deployed on limited and decentralized computing infrastructures.
In this work, we study the emerging paradigm of stateful Function as a Service (FaaS) with lightweight task abstractions in WebAssembly.
Specifically, we assess the implications of deploying inter-dependent tasks with an internal state on edge computing resources using a stateless vs. stateful approach and then derive a mathematical model to estimate the energy consumption of a workload with given characteristics, considering the power used for both processing and communication.
The model is used in extensive simulations to determine the impact of key factors and assess the energy trade-offs of stateless vs. stateful.
\end{abstract}

}

\maketitle

\begin{tikzpicture}[remember picture,overlay]
\node[anchor=south,yshift=10pt] at (current page.south) {\fbox{\parbox{\dimexpr\textwidth-\fboxsep-\fboxrule\relax}{
  \footnotesize{
     \copyright 2024 IEEE.  Personal use of this material is permitted.  Permission from IEEE must be obtained for all other uses, in any current or future media, including reprinting/republishing this material for advertising or promotional purposes, creating new collective works, for resale or redistribution to servers or lists, or reuse of any copyrighted component of this work in other works.
  }
}}};
\end{tikzpicture}%

\IEEEdisplaynontitleabstractindextext

\IEEEpeerreviewmaketitle


\section{Introduction}\label{sec:introduction}

Since long edge computing has outgrown the role of being a surrogate of the cloud for offloading specific services it was originally assigned (e.g.,~\cite{Chang2014}).
Nowadays, it is a thriving reality with a transformative effect on business and society, whose adoption is constantly spreading across more and more domains~\cite{ramalingam_leading_2023}, especially thanks to the opportunities offered by cognitive \ac{AI} services at the edge, which can benefit immensely from the data being consumed close to where they are generated and from the use of decentralized computing resources with embedded \acp{GPU}/\acp{TPU}~\cite{otto_edge_2023}.

\begin{figure}[tb]
\centering
\includegraphics[width=\columnwidth]{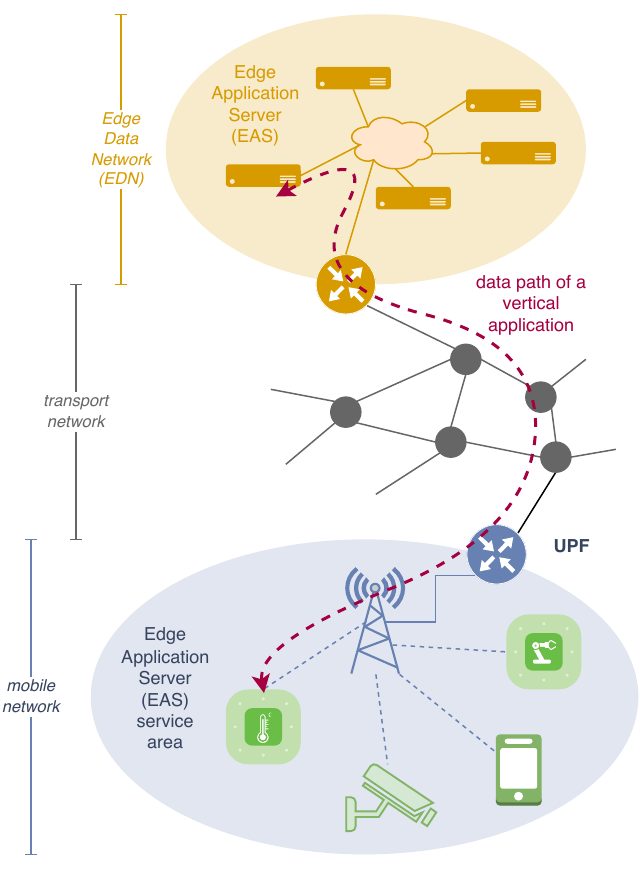}
\caption{Simplified beyond 5G architecture of an \acf{EDN} mobile users with \acf{EAS} services.}
\label{fig:intro}
\end{figure}

Indeed, edge computing is expected to remain a key player in the future evolution of communication technologies~\cite{wang_road_2023} and has attracted the interest of many standardization bodies, including 3GPP. In particular, a reference architecture for supporting edge computing within a cellular network is standardized in\cite{noauthor_3gpp_2023}, enabling \acp{ECSP} to deploy \ac{EDN}s, namely 5G network infrastructures that contain edge servers (called EAS in 3GPP jargon) and management entities, which can offer edge computing services to end users. A typical deployment scenario for an EDN is shown~\Cref{fig:intro}, together with the interactions between the 5G transport functions and the edge computing entities. At the bottom layer of this reference architecture, there is the mobile network with \ac{IoT} users, possibly consisting of a network slice~\cite{foukas_network_2017}, that is a virtual network dedicated to a customer in logical isolation, and providing specific \acp{SLA}, while sharing the physical \ac{RAN} and \ac{5GC} resources\cite{zeydan_service_2022}.
Virtual applications can be assigned by the customer to a given \ac{EAS} service area of the mobile network, whose traffic is handled by one \ac{UPF} (there can be more than one, not shown in the figure) towards a transport network until it reaches the \ac{EDN}, at the top of the diagram.
The \ac{EDN} is a pool of computing, networking, and storage resources contributing to the \ac{EAS}.

From the point of view of software architectures, cloud-native approaches based on micro-services and containers, instead of monolithic applications, have become very popular,  especially relying on serverless computing.
The latter enables full automation and flexible billing~\cite{kounev_serverless_2023}, building on a functional programming model, called \ac{FaaS}, which allows a developer to compose elementary and reusable \emph{functions}, whose instances are stateless and can be shared among different users.
Serverless computing has been adopted successfully in many application domains in the cloud~\cite{eismann_state_2021}, but it is not yet fully developed at the edge~\cite{raith_serverless_2023}, because of two main reasons.
First, real applications often require functions to maintain a state, of some kind, associated with a given user or session.
In the cloud, this is implemented through storage services, which are readily available in data centers, but incur much more overhead in a decentralized edge computing environment.
Second, efficient scheduling and resource management are much more challenging at the edge due to the scarcity of resources and the resulting contention between concurrent applications, which hampers the promises of full automation and infinite scalability of serverless computing.

These observations lead to the following research questions:

\begin{description}
\item[Q1] \emph{Deploying stateful FaaS through stateless runners accessing the state on an external service is the state-of-the-art deployment option. Is there an alternative solution to obtain the benefits of edge computing and the flexibility of FaaS, despite the need to maintain a state?} 

\item[Q2] \emph{Improving the efficiency of \ac{ICT} is one of the key goals for global sustainable development~\cite{freitag_real_2021}. What are the implications of different deployment/run-time options to support stateful FaaS on the energy consumption of the \ac{EDN} infrastructure?}
\end{description}

We address both questions through the paper contributions:

\begin{description}
\item[A1] We consider an alternative paradigm that realizes \emph{stateful \ac{FaaS}} (also known as ``actor model'' in the literature, e.g., \cite{de_boer_survey_2024}) exploiting recent lightweight virtualization technologies and allowing efficient multiplexing of many instances of isolated user-space applications. This is explored in \Cref{sec:motivation} below, where we provide a background on the deployment of inter-dependent stateful tasks through stateless vs. stateful FaaS and report the results obtained in a testbed with embedded edge nodes on the execution of concurrent tasks in WebAssembly.
\item[A2] We formulate in \Cref{sec:model} a novel mathematical model for the energy consumption of \acp{EDN} using the two stateless vs. stateful FaaS approaches above, which is then used to carry out a performance evaluation study under realistic trace-inspired workload conditions and analyze the impact on energy consumption of the main system factors (\Cref{sec:eval}).
\end{description}

The paper continues with a review of the relevant state-of-the-art in \Cref{sec:soa}, and then comes to the summary\optional{and future work} in \Cref{sec:conclusions}.

\section{Stateless vs. Stateful FaaS:\\Background and Motivation}\label{sec:motivation}

Serverless computing and the \ac{FaaS} programming model are popular in the cloud~\cite{eismann_state_2021} and they have attracted significant interest also at the edge~\cite{risco_rescheduling_2024}.
As already mentioned in \Cref{sec:introduction}, with \ac{FaaS} an application is made of a sequence of stateless function calls, which can be arranged in chains (i.e., $f_1 \rightarrow f_2 \rightarrow \ldots \rightarrow f_N$) or more complex structures, like \acp{DAG}~\cite{mahgoub_wisefuse_2022}.
However, realistic applications typically \emph{do need} function execution to be associated with some state, especially for edge applications, such as \ac{AI} and real-time analytics~\cite{xu_stateful_2023}.

\begin{figure}[tb]
\centering
\includegraphics[scale=0.8]{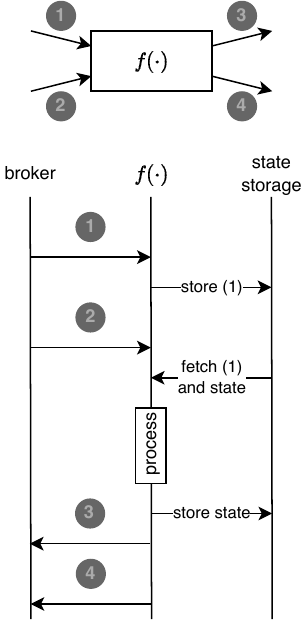}
\caption{Example of how to realize stateful processing with stateless FaaS.}
\label{fig:stateless-seq}
\end{figure}

A straightforward solution to this problem, which we call \textbf{stateless FaaS}, is to maintain the state on an external storage system to be accessed on demand by the functions as part of their execution, as explained, e.g., in~\cite{Xie2021}.
Such a deployment option is illustrated in the example in \Cref{fig:stateless-seq}, where function $f(\cdot)$ requires input from two dependencies (1 and 2) and has two outputs (3 and 4).
When the function receives input 1, it is kept temporarily in the state storage.
Once input 2 is received, full processing can occur combining the latter with the previous input 1 and the state, to produce the final outputs 3 and 4, after updating the state on the storage.

\begin{figure}[tb]
\centering
\includegraphics[scale=0.8]{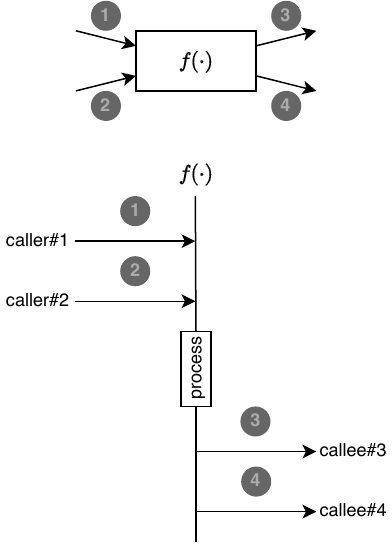}
\caption{Example of stateful FaaS.}
\label{fig:stateful-seq}
\end{figure}

In common serverless computing platforms, function invocation happens through an HTTP command issued on a web server running in a container.
Due to the lack of state, the same container can serve multiple users/sessions seamlessly, and the orchestration platform can easily perform autoscaling of such runners, i.e., decreasing or increasing the number of instances per function to match the instantaneous demand.

\begin{figure}[tb]
\centering
\includegraphics[scale=0.8]{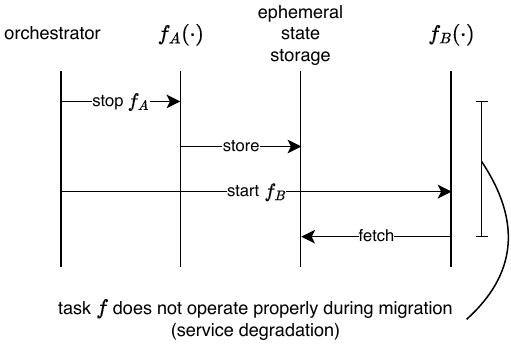}
\caption{Migration of a stateful FaaS runner from node $A$ to node $B$.}
\label{fig:stateful-migration}
\end{figure}

An alternative to this strategy is dedicating each user/session to a runner, thus realizing what we call \textbf{stateful FaaS}.
As illustrated in the example in \Cref{fig:stateful-seq}, with this model there is no need to fetch/update the state or store temporary input from previous function calls.
In principle, the stateful FaaS model has two inconveniences.
First, the number of runners may be much higher than that with stateless FaaS, because the former cannot exploit statistical multiplexing of multiple users/sessions like the latter.
We study the practical impact of this issue with a specific virtualization technology in \Cref{sec:motivation:experiments}.
Second, if a runner is migrated from one node to another for any reason, e.g., system resource optimization, its internal state must be moved to the target host.
We show an example in \Cref{fig:stateful-migration}, where the orchestrator migrates a runner for the function $f(\cdot)$ from node $A$ to node $B$.
First, when stopping $f(\cdot)$ on node $A$ the state is stored temporarily on an external system, which is then queried by the new instance of function $f(\cdot)$ on node $B$ upon creation.
With this solution, there would be a period during which the task performed by $f(\cdot)$ is not available.
More sophisticated protocols can be devised~\cite{cicconetti_faas_2022}, but, in any case, they would incur additional complexity or overhead, which is not needed with stateless FaaS.
The impact of state migration on energy consumption is captured by the mathematical model defined in \Cref{sec:model} and evaluated in \Cref{sec:eval}.

\begin{figure}[tb]
\centering
\includegraphics[scale=0.8]{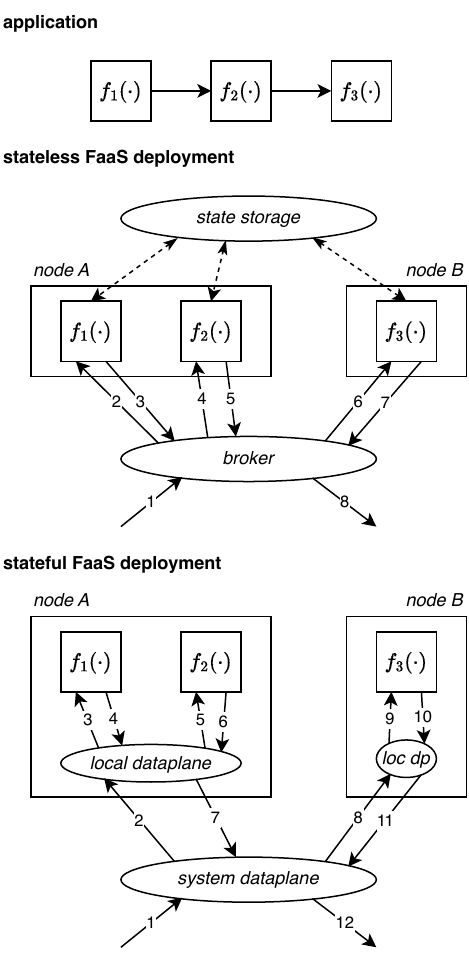}
\caption{Deployment of a three-function chain (top) on two processing nodes through stateless FaaS (middle) and stateful FaaS (bottom).}
\label{fig:stateless-vs-stateful}
\end{figure}

We now illustrate deployment with stateless vs. stateful FaaS with the help of the example in \Cref{fig:stateless-vs-stateful}, with a three-function chain application running on two nodes A and B.
In the example, we have one runner per function: node A hosts functions $f_1$ and $f_2$ and node B hosts function $f_3$.
With stateless FaaS, an intermediate layer is needed to dispatch function invocations to one of the matching runners: this is represented by a logical component called \emph{broker}, borrowing the terminology from~\cite{schafer_tasklets_2016}, which is an early study on the realization of distributed computing in pervasive systems.
As can be seen, network traffic is generated at each function call for state access, on the state storage, and for invoking the next runner through the broker.
On the other hand, with stateful FaaS, we need logical components to mesh together the runners, which can be within a node or at a system level.
Network access for accessing the state is unnecessary because the state is embedded within the runner.
Furthermore, when a runner invokes another on the same node no network access is needed, too. 

\subsection{Stateful FaaS benchmarking}\label{sec:motivation:experiments}

We now discuss the potential issue of stateful FaaS requiring one runner per user/session, which can be substantially mitigated with lightweight abstractions.
A candidate technology for this is WebAssembly, which is new but whose adoption is increasing very fast.
With WebAssembly the runner is a user-space application, whose isolation is provided by the run-time environment at a small fraction of the cost of traditional virtualization means, such as Docker containers~\cite{gackstatter_pushing_2022}.

\begin{figure*}[tb]
\centering
\includegraphics[scale=0.8]{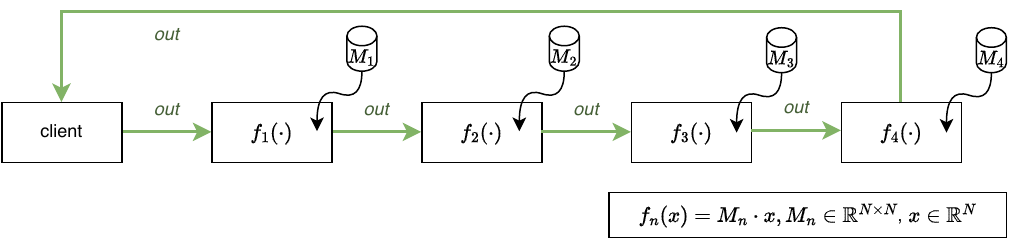}
\caption{WebAssembly stateful FaaS: app consisting of four functions in a chain.}
\label{fig:motivation-workflow-2}
\end{figure*}

In the remainder of this section, we report experimental results obtained by running an application with four stateful functions in a chain, as illustrated in~\Cref{fig:motivation-workflow-2}.
The implementation was done with EDGELESS\footnote{\url{https://github.com/edgeless-project/edgeless}}, an open-source framework to develop and orchestrate stateful FaaS applications in the edge-cloud continuum, developed within a project funded by the European Union with the same name.
The state of each function is a square matrix of real values and the processing consists of multiplying it by the input argument and providing the result as output.
The size $N$ of the input/output vector, and hence the matrix, is 500 and all the states and inputs are initialized with random values in $[0,1]$.
The chain is triggered by a client every time it receives the final output of the chain execution on the previous round, thus each app generates a rate of invocation as high as allowed by the underlying platform/hardware.
We ran the experiments on two types of embedded devices: a Raspberry Pi 3 and an NVIDIA Jetson NX.
Each experiment lasted 10 seconds after an initial warm-up period (of variable size) and was repeated 10 times.
The error bars report low (0.025) and high (0.975) quantiles of measured data.

\begin{figure}[tb]
\centering
\includegraphics[width=\columnwidth]{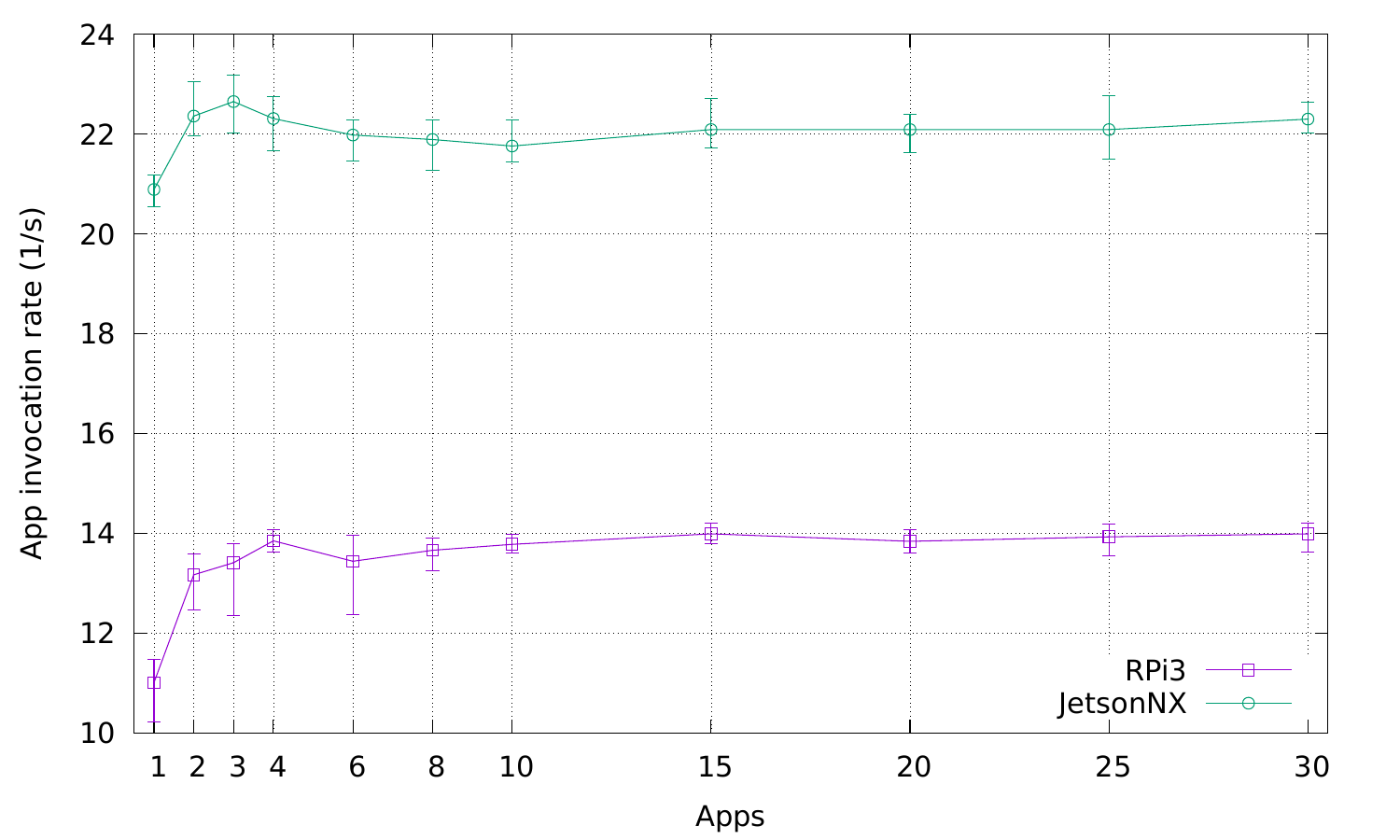}
\caption{WebAssembly stateful FaaS: app invocation rate vs. number of apps.}
\label{fig:motivation:tpt}
\end{figure}

In \Cref{fig:motivation:tpt} we show the invocation rate with an increasing number of apps.
The NVIDIA Jetson NX curve lies above the Raspberry~Pi~3 because it has a more powerful CPU.
With both hardware, the curve increases from 1 app to 2 apps, because the former is not sufficient to saturate the CPUs on the host node.
However, the invocation rate afterward remains stable, even with 30 apps, corresponding to 120 concurrent runners (see \Cref{fig:motivation-workflow-2}) in the same embedded device, with no noticeable drop in aggregated performance.
This confirms that stateful FaaS runners with WebAssembly can scale with negligible overhead until high load levels, relative to the node capabilities.
This assumption is used in the next section.



\section{System Model}\label{sec:model}

We now define a mathematical model to estimate the energy consumed in a time horizon $T$ for executing the applications that enter/leave the system during that period.
The model is intended to be used to evaluate high-level deployment strategies and run-time orchestration policies and, as such, it is not intended to provide quantitatively accurate results, but rather qualitative guidelines to drive algorithm design and high-level resource provisioning.

\begin{figure}[tb]
\centering
\includegraphics[scale=1]{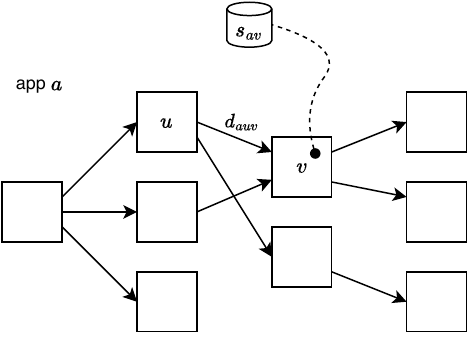}
\caption{Application model. An app $a$ consists of functions arranged in a graph. If function $u$ calls function $v$ then an edge exists, and its weight $d_{auv}$ is the amount of data exchanged. Each function $v$ has a state of size $s_{av}$.}
\label{fig:app}
\end{figure}

We assume the workload is made of applications (apps for short) that enter and leave the system dynamically at given times $t^\downarrow_a$ and $t^\uparrow_a$, for app $a$.
An app $a$ consists of some functions (or tasks) arranged in a directed dependency graph $G_a(V_a,E_a)$.
Each vertex $v \in V_a$ is a task that depends on its predecessors (incoming edges) and produces output towards its successors (outgoing edges).
The amount of data exchanged when task $u$ calls its successor task $v$ is $d_{auv}$, in bits.
Without loss of generality, to have a more compact notation, we assume that the invocation rate is common for all the tasks within app $a$ and equal to $\lambda_a$.
Task $v$ has an internal state of size $s_{av}$, in bits, and a processing request equal to $r_{av}$, in fractions of CPU.
An example of a dependency graph is illustrated in \Cref{fig:app}.

In the following, we consider the system as dynamic, characterized by a series of discrete events happening at time $t_k \in \{t1, \ldots, t_N\}$, where $t_N$ is the end of the period of interest and the other events correspond to an application entering or leaving the system.
Between two consecutive events the power consumption remains stable (in a statistical sense) and we can characterize its average value through two step-wise functions, which are constant from time $t_k$ until the next event $t_{k+1}$: $\alpha(t_k)$ is number of edge nodes used at time $t_k$ to serve the active applications, where each node has a processing capacity $C$, in fractions of CPU; $\beta_a(t_k)$ is the average network traffic consumed by application $a$ in the unit of time.
We assume that the power consumption of an edge node is binary: if it is used, i.e., it serves at least one stateless FaaS or hosts at least one stateful FaaS runner, then it consumes a peak power; otherwise, if it is unused, it does not consume power at all.
Such an assumption does not capture the features offered by state-of-the-art power management technologies, which allow nodes to be in multiple states (active, idle, sleep, etc.) and to scale the frequency of individual CPU cores, but contributes to keeping the model simpler and compact.
In our future work, we will expand to more sophisticated models.

\begin{table}[tbp]
\caption{Notation used in the paper. The last two rows are used only with Stateful FaaS.}
\centering
\begin{tabularx}{\columnwidth}{c|X|c}
    \thead{Parameter} & \thead{Description} & \thead{Unit} \\
    \hline
    $G_a(V_a,E_a)$ & Task graph of app $a$. $V_a$ is the set of tasks, $E_a$ represents the invocation dependencies & \\
    $\lambda_a$ & Invocation rate of app $a$ & s$^{-1}$ \\
    $r_{av}$ & Processing request of task $v$ at app $a$ & CPU \\
    $s_{av}$ & State size of task $v$ at app $a$ & b \\
    $d_{auv}$ & Invocation data size from task $u$ to $v$ at app $a$ & b \\
    $t^{\downarrow}_a$ & Arrival time of app $a$ & s \\
    $t^{\uparrow}_a$ & Leaving time of app $a$ & s \\
    $A$ & Set of all the applications in the period of interest & \\
    $A(t_k)$ & Set of applications active at time $t_k$ & \\
    \hline
    $\alpha(t_k)$ & Number of edge nodes active at time $t_k$ & \\
    $\beta_a(t_k)$ & Traffic rate of app $a$ at time $t_k$ & b/s \\
    $P_N$ & Power consumption of a node & W \\
    $E_B$ & Per-bit network transfer energy & J/b \\
    $E$ & Total energy consumed in the period of interest & J \\
    $C$ & Processing capacity of a node & CPU \\
    $t_k$ & Time instant of the $k$-th event & s \\
    \hline
    $\Delta$ & Defragmentation interval & s \\
    $x_{av}(t_k)$ & Mapping function indicating the index of the node to which task $v$ of app $a$ is allocated at time $t_k$ & \\
    \hline
\end{tabularx}
\label{tab:notation}
\end{table}

Regardless of the deployment strategy, we can then define the total energy consumed in the system as follows:

\begin{equation}\label{eq:energy}
\begin{aligned}
E & = \sum_{k=1}^{N-1} \bigg[ P_N \cdot \alpha(t_k) \\
  & + E_B  \sum_{a \in A} \beta_a(t_k)\mathbb{I}(t^\downarrow_a \leq t_k \leq t^\uparrow_a)
    \bigg] (t_{k+1}-t_k),
\end{aligned}
\end{equation}
where $P_N$ is the power consumption of an edge node and $E_B$ is the per-bit network transfer energy, and $\mathbb{I}(\cdot) \in \{0,1\}$ is an indicator function equal to 1 if and only if the condition is true, which in \Cref{eq:energy} means that the app $a$ is active at time $t_k$.
In this work, we focus on energy consumption assuming that there are no constraints on the availability of processing and network resources.
In other words, we assume that the system can accommodate all the incoming requests, hence no admission control is needed.
The notation used in the paper is summarized in \Cref{tab:notation}.

In the following we define $\alpha(t_k)$ and $\beta_a(t_k)$ separately for stateless FaaS (\Cref{sec:model:less}) and stateful FaaS (\Cref{sec:model:full}), based on the considerations in \Cref{sec:motivation} above.

\subsection{Stateless FaaS}\label{sec:model:less}

For stateless FaaS we adopt a simple model that captures well its distinguishing features.
Specifically, we assume that the number of active nodes needed at time $t_k$ is the minimum possible, i.e.:

\begin{equation}\label{eq:less:alpha}
    \alpha(t_k) = \left\lceil
    \frac{1}{C} 
    \sum_{a \in A(t_k)} 
        \sum_{v \in V_a} r_{av}
    \right\rceil,
\end{equation}
where $A(t_k)$ is set of applications active at time $t_k$.
The inner summation in \Cref{eq:less:alpha} is the total processing request of app $a$, which is then summed over all the applications and, finally, divided by the edge node capacity $C$.
This implicitly assumes that no edge effects exist in horizontal scalability and the broker layer can distribute the load appropriately among the multiple task instances. 
On the other hand, the traffic rate of app $a$ at time $t_k$ is given by:

\begin{equation}
\beta_a(t_k) = \lambda_a \left[ \sum_{v \in V_a} s_{av}
    + \sum_{(u,v) \in E_a}d_{uav}\right],
\end{equation}
which is the sum of the traffic generated for the state access (first term) and function invocation between each node and its successors (second term), in the unit of time, as given by the invocation rate $\lambda_a$.

\subsection{Stateful FaaS}\label{sec:model:full}

The model with stateful FaaS is more complicated because it depends on how tasks are assigned to edge nodes for three reasons.
First, function invocation only consumes network resources if the two tasks are not assigned to the same edge (see \Cref{fig:stateless-vs-stateful}).
Second, since a stateful FaaS runner cannot be split/recombined, assigning the active tasks to available nodes to minimize the number of nodes used is akin to the bin-packing problem, which is known to be NP-complete~\cite{garey_computers_2009}.
Finally, as active apps leave the system, \emph{fragmentation} occurs (a term inspired by the similar effect in the memory management process of operating systems), i.e., edge nodes are only partially allocated: this is sub-optimal for energy consumption.
To solve this problem, we foresee a defragmentation process to happen periodically, with the period equal to $\Delta$, which is a system configuration parameter: during defragmentation, the active apps are rearranged to reduce the number of edge nodes needed, thus saving energy in the future.
However, this process \emph{consumes} energy because the state of some runners may have to be migrated from one node to another (see \Cref{fig:stateful-migration}).
In \Cref{sec:eval} we will study the trade-off in choosing the value of $\Delta$.



Now we introduce a last bit of notation: let $x_{av}(t_k)$ be a variable that indicates what edge node (using an arbitrary indexing scheme) hosts the runner for the task $v$ of app $a$ at time $t_k$.
In time intervals where the app $a$ is inactive, i.e., before it enters or after it leaves the system, the variable is undefined.
The values of $x_{av}(t_k)$ must be determined through two orchestration decision-making algorithms: i) when an app enters the system, the algorithm chooses where to deploy each of its tasks, by either selecting edge nodes already active (hosting other tasks) with sufficient residual capacity or activating new edge nodes; ii) upon defragmentation, the tasks of active applications can be migrated to other edge nodes to reduce the total number of the active ones.
Determining an optimal policy for either of these decision processes has the same complexity as finding an optimal allocation for a bin-packing problem, as already mentioned.
To keep the focus of this work compact, we defer the study of those problems to future work and we propose to use the following simple heuristic based on the \emph{best-fit} policy, which is widely employed and has bounded performance~\cite{martello_bound_1981}.

\noindent\ul{Stateful\textbar best-fit algorithm:}
\begin{itemize}[parsep=0em,leftmargin=*,label={--}]
    \item When an app enters the system, for each task we select the active node that hosts one of the predecessor tasks, if any (to save network traffic for function invocation). Otherwise, we select the active node that leaves the smallest residual capacity, if any, breaking ties arbitrarily. Otherwise, we deploy the task on an inactive node.
    \item Upon defragmentation, we apply the above algorithm policy to all the active apps, in arbitrary order.
\end{itemize}%

We then derive the number of active nodes at time $t_k$ as:

\begin{equation}\label{eq:full:alpha}
\alpha(t_k) =
    \Big|
    \big\{
    x_{av}(t_k), \forall a \in A(t_k), \forall v \in V_a
    \big\}
    \Big|,
\end{equation}
where $|\cdot|$ indicates the cardinality of the corresponding set, and the traffic rate of app $a$ at time $t_k$ is:

\begin{equation}\label{eq:full:beta}
\begin{aligned}
\beta_a(t_k) = &
    \frac{1}{t_{k+1}-t_k} \sum_{v \in V_a} s_{av} \cdot \mathbb{I}\left(
        x_{av}(t_k) \neq x_{av}(t_{k-1})
        \right) + \\
    & \lambda_a \sum_{(u,v) \in E_a} d_{auv} \cdot \mathbb{I}\left(
        x_{au}(t_k) \neq x_{av}(t_k)
        \right),
\end{aligned}
\end{equation}
where the first addend takes into account the state migration if the task was moved since the previous time event (by design, this can happen only during the defragmentation procedure) and the second addend considers the network traffic for function invocation, only if the task $u$ and its successor $v$ do not belong to the same node (see \Cref{fig:stateless-vs-stateful}, again).

\section{Performance Evaluation}\label{sec:eval}

In this section, we evaluate the performance, in terms of energy consumption, of the stateless vs. stateful approaches, as modeled in \Cref{sec:model}, indicated as \emph{stateless\textbar min-nodes} and \emph{stateful\textbar best-fit}, respectively.
For reference purposes, we also include two alternatives: \emph{stateless\textbar max-balancing}, as implied by the name, refers to a stateless FaaS system that seeks to maximize load balancing~\cite{cicconetti_decentralized_2020}; \emph{stateful\textbar random} is a variation of the stateful policy in \Cref{sec:model:full}, where there is no periodic defragmentation and the tasks of incoming apps are assigned to edge nodes at random, respecting the maximum capacity $C$, and a new node is made active only if there is none with sufficient residual capacity.
For full reproducibility of results, the source code of the simulator and the scripts and artifacts are available publicly as open-source on GitHub\footnote{\url{https://github.com/ccicconetti/stateful-faas-sim} (experiment \texttt{001})}.

\noindent\textbf{Methodology and assumptions} The workload is created following the model in~\cite{Tian2019}, which is inspired by real traces made available by Alibaba and broadly used in the literature, tuned as follows: the arrival and lifetime of apps follow a Poisson distribution, with average 1~s and 60~s, respectively; both the state size $s_{av}$ and the data invocation size $d_{auv}$ are derived from the memory requirements produced by~\cite{Tian2019}, by applying multiplicative factors called $S$ (state) and $D$ (data invocation), where $D$ is always set to 100, which corresponds to the range $[2,303]$~kB, and $S$ is expressed through the ratio $S/D$, which is 100 by default, in which case $S$ would be in the range $[0.2,30.3]$~MB.
The invocation rate is $\lambda_a=$~5/s and the capacity of a node $C$ is set to 1000, which is sufficient to host any single task, whose requested capacity is drawn from an empiric distribution with a maximum value of 800.
The edge node power consumption was set to 100~W, which is typical for a small device such as an Intel NUC; estimating the network consumption is much more complicated because it depends not only on the devices but also on the overall networking infrastructure: based on the results from a recent study~\cite{ahvar_estimating_2022}, we have experimented with different values in the range $[0.05,5]$~$\mu$W/b/s.
Each experiment lasted 1~day of simulated time and was repeated 1000~times; the plots show the average value across the repetitions with a symbol and the low (0.025) and high (0.975) quantiles as error bars.
All the values above are to be considered unless specified otherwise.

\begin{figure}[tb]
\centering
\includegraphics[width=\columnwidth]{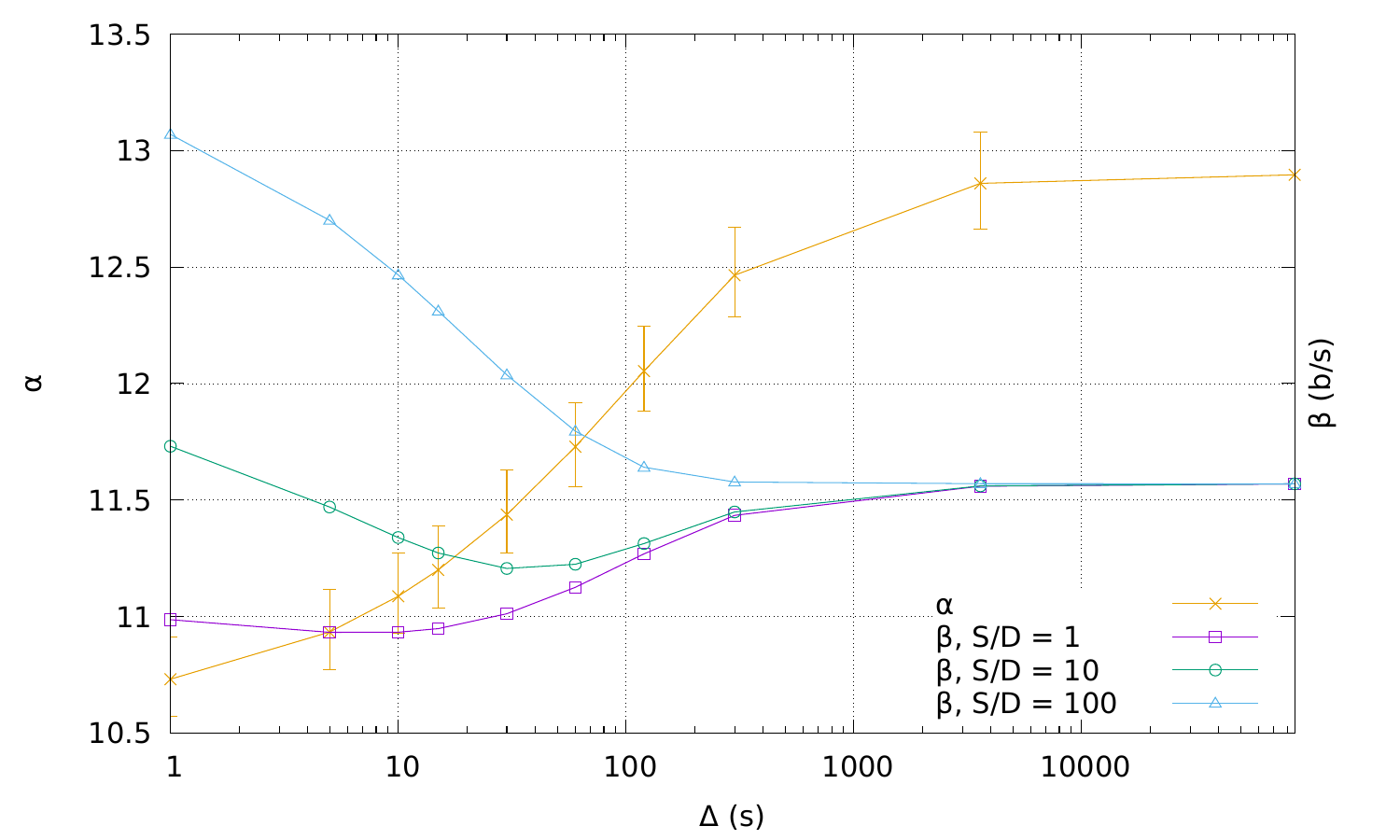}
\caption{Simulations: $\alpha$ and $\beta$ vs. defragmentation period $\Delta$.}
\label{fig:001-var-defrag}
\end{figure}


\noindent\textbf{Defragmentation period}
In \Cref{fig:001-var-defrag} we show $\alpha$ and $\beta$ with different combinations of $\Delta$ and the $S/D$ ratio, only with stateful\textbar best-fit.
$\beta$ is affected significantly by both $S/D$ and $\Delta$: when the state is heavier ($S/D=100$), the network traffic is very high with small values of $\Delta$ (note the log scale on the $y$-axis) because frequent migrations are expensive.
This effect is much less prominent with $S/D=10$ and $S/D=1$, because of the smaller state sizes compared to the invocation data sizes.
With increasing $\Delta$, all the curves initially decrease and, then, increase again until they converge to the same value (as the defragmentation becomes more sporadic, the state size becomes less important).
The minima of the curves depend on the specific value of $S/D$.
The number of active nodes $\alpha$ is independent of $S/D$ and always increases with $\Delta$.
\underline{Key message:} \textit{The choice of $\Delta$ incurs a trade-off in the energy consumption of computation vs. network.}
Devising an algorithm to set $\Delta$ at run-time is a possible spin-off research activity.
In the following, we set the value to 120~s, i.e., twice the average app lifetime, which appears as a reasonable trade-off between network vs. processing consumption.

\begin{figure}[tb]
\centering
\includegraphics[width=\columnwidth]{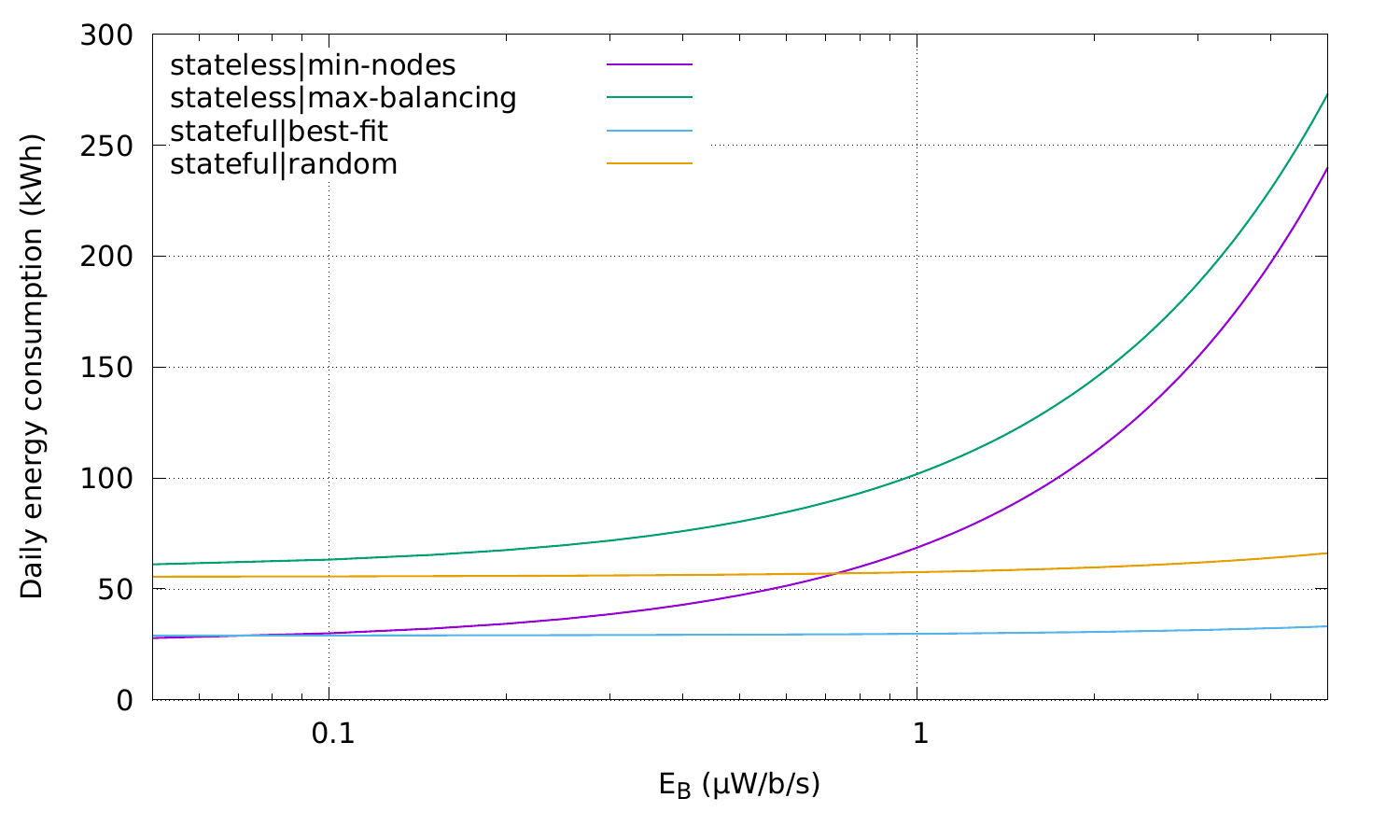}
\caption{Simulations: energy consumption vs. $E_B$.}
\label{fig:001-var-energy}
\end{figure}

\noindent\textbf{Energy per bit-rate}
In \Cref{fig:001-var-energy} we show the energy consumption with increasing $E_B$ while keeping $P_N=$~100~W.
\underline{Key message:} \textit{The energy consumption increase with a higher per-bit-rate cost is higher with a stateless deployment, especially in the max-balancing flavor, and is very modest with a stateful deployment.}
In the latter case, we can see that the best-fit policy reduces energy consumption by about 2 compared to random, for all values of $E_B$.
In the following, we only consider the two extremes of the $E_B$ range.

\begin{figure}[tb]
\centering
\includegraphics[width=\columnwidth]{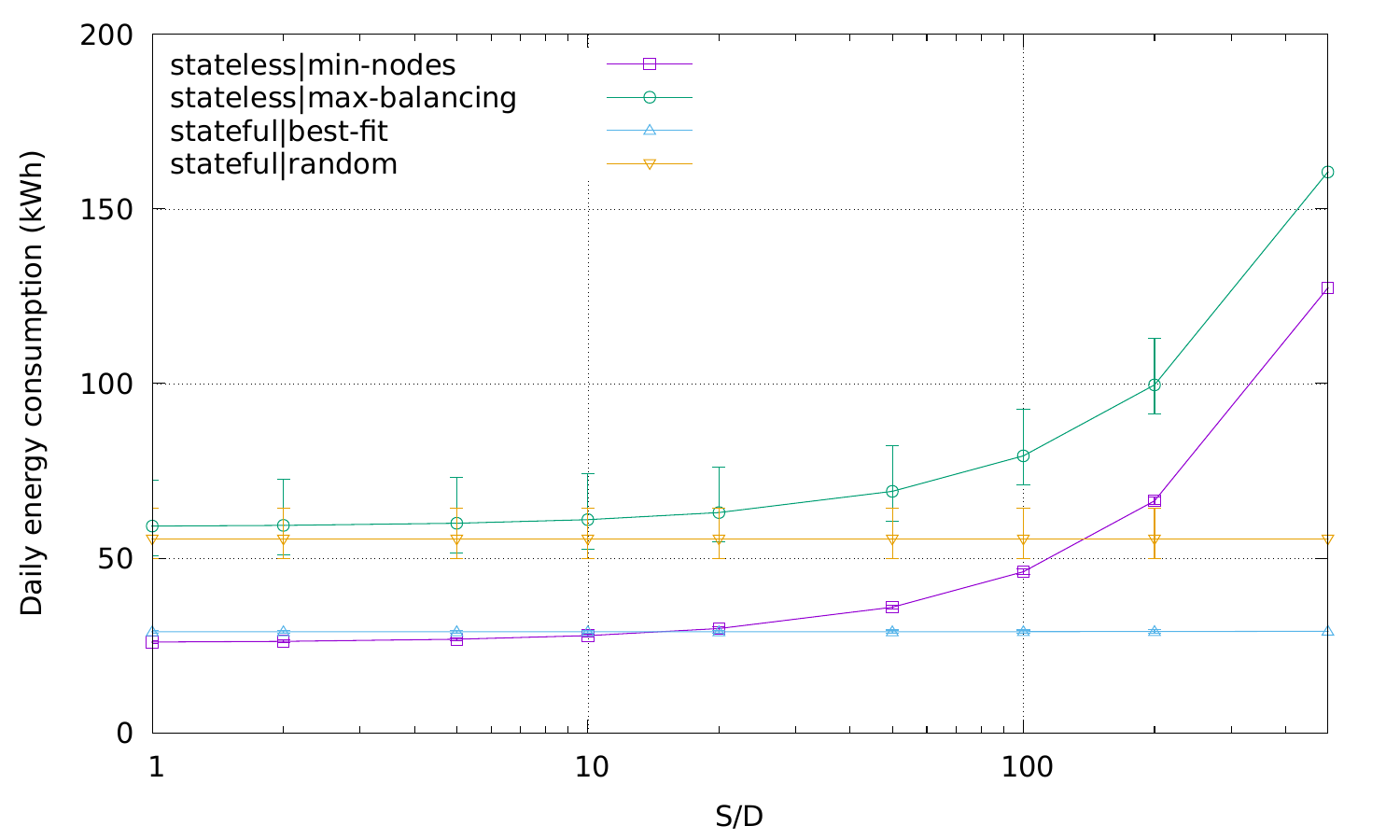}
\caption{Simulations: energy consumption vs. $S/D$, $E_B=0.05$~$\mu$W/b/s.}
\label{fig:001-var-state}
\end{figure}

\noindent\textbf{State size}
The impact of the state size, compared to the data invocation size, is exposed in \Cref{fig:001-var-state}, with low per-bit-rate energy cost, i.e., $E_B=0.05$~$\mu$W/b/s.
A stateless deployment, with a min-nodes policy, is the best option only for $S/D \leq 10$ and only by a small margin compared to stateful\textbar best-fit.
On the other hand, as $S/D$ increases significantly above 10, stateless deployment becomes significantly more energy-hungry, due to the cost of accessing the state upon each function invocation.
With $S/D > 100$, stateless is outperformed even by stateful\textbar random.
The max-balancing policy follows the same trend as min-nodes and is always above the latter, though the gap reduces slightly as $S/D$ increases.
\underline{Key message:} \textit{From an energy consumption perspective, stateful deployments are almost insensitive to the size of the applications' states.}

\begin{figure}[tb]
\centering
\includegraphics[width=\columnwidth]{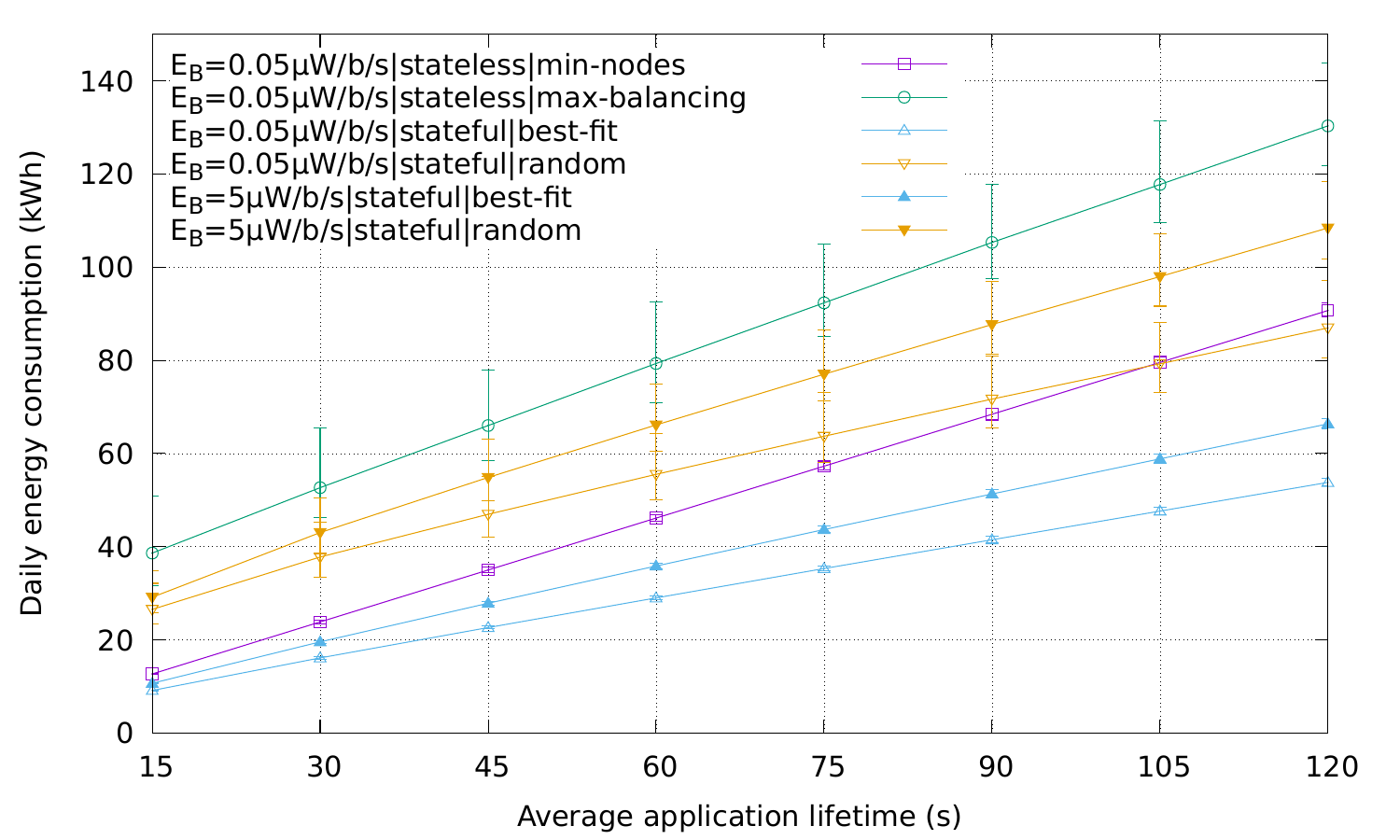}
\caption{Simulations: energy consumption vs. average application lifetime.}
\label{fig:001-var-load}
\end{figure}

\noindent\textbf{System load}
In \Cref{fig:001-var-load} we report the measurements obtained with min/max $E_B$ values for stateful policies (with stateless, the values with maximum $E_B$ are well above the plot $y$-axis range) when increasing the application lifetime from 15~s to 120~s.
As expected, all the curves increase with the load.
\underline{Key message:} \textit{Both stateful\textbar best-fit curves lie at the bottom and gain an increasing margin compared to all the others as the load increases.}
The second-best option is stateless\textbar min-nodes (only with minimum $E_B$), while the stateless\textbar max-balancing performs worst.

\begin{figure}[tb]
\centering
\includegraphics[width=\columnwidth]{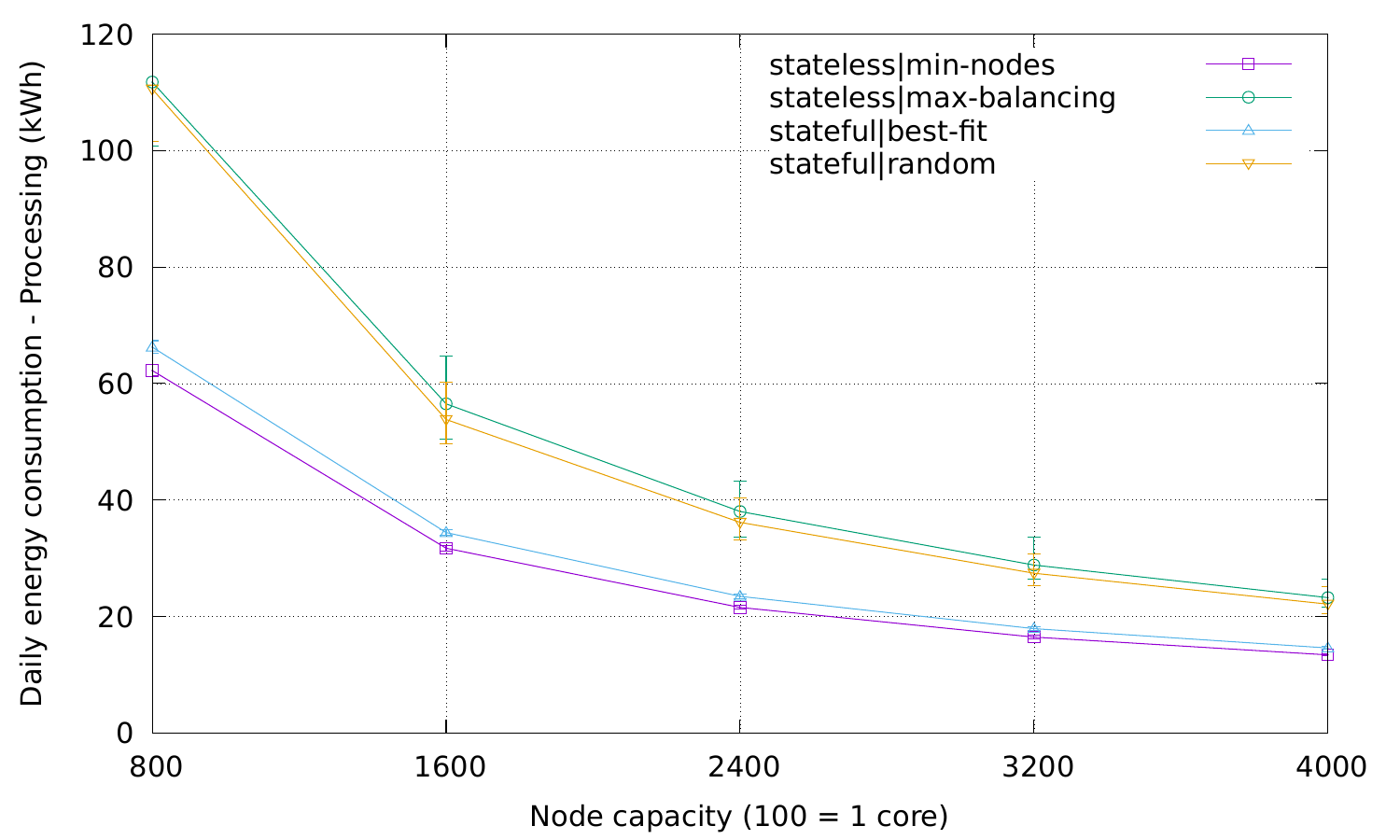}
\caption{Simulations: energy consumption vs. node capacity.}
\label{fig:001-var-capacity}
\end{figure}

\noindent\textbf{Node capacity}
Finally, in \Cref{fig:001-var-capacity} we show the energy consumption (only due to processing) with increasing $C$ from 800 to 4000.
All the curves decrease because with increasing $C$ the number of nodes required decreases, as well, while we keep the power consumption per node $P_N$ constant.
It is interesting to note that the curves are almost overlapping in pairs.
At the bottom (less energy consumed) we find the two systems described in \Cref{sec:model}: in fact, they both aim at reducing the edge computing infrastructure energy consumption.
Stateless has a slight gain compared to stateful, but it is more than compensated by a lower energy efficiency from the network traffic perspective.
At the top (more energy consumed), the two comparison systems show similar performance, which can be explained by the fact that they both try to spread as much as possible the load among the active nodes: stateless\textbar max-balancing does this explicitly, stateful\textbar random implicitly.
\underline{Key message:} \textit{A stateful deployment, with a best-fit allocation strategy, can be as efficient as a stateless one despite the fragmentation issue.} 

\section{Related Work}\label{sec:soa}



\subsection{Edge computing support in 5G systems}\label{sec:soa-edn}

The Multi-access Cloud Computing (MEC) concept was defined by ETSI~\cite{etsi_mec_2022} to offer a standardized framework to include cloud-computing capabilities to edge networking elements, such as base stations, access points, switches, and routers.
Several papers have investigated system architecture and technology enablers to build MEC-enabled shared pools of computing resources at the network edge, especially within cellular networks~\cite{han_mec_2018}.
More recently, the 3GPP has extended the reference design of MEC systems to support the deployment of edge computing applications within 5G and post-5G networks~\cite{3gpp_edn_2023}, by defining the underlying layer that facilitates communication between application clients (ACs) running on the user terminal and application servers (EAS) hosted by an EDN.
Furthermore, the 3GPP technical specifications delineate the capabilities for edge server discovery, configuration, and management, encompassing service continuity and the exposure of networking properties.
In addition to standardization activities, research studies have investigated possible reference architectures for EDN.
For instance, the problem of optimal placement of edge servers in a 5G network is addressed in~\cite{zhang_planning_2023} to maximize the total revenues or to minimize the deployment costs throughout a planning horizon~\cite{wei_planning_2022}.
In~\cite{zhang_sharing_edge_2020}, the authors argue that an EDN can improve the utilization of computing resources by cooperating with a regional cloud through a wholesale and buyback pricing scheme.

\subsection{Application deployment} \label{sec:soa-app}

The authors in the recent survey~\cite{hazra_meeting_2023} identified resource allocation and scheduling as an open research challenge to unlock the full potential of \ac{IoT} applications at the edge. In~\cite{xu_stateful_2023}, the authors argue that most \ac{AI} applications are stateful and propose an \ac{ILP} to minimize the deployment cost, defined as a combination of processing, transmission, and storage, which then solve via a heuristic algorithm with provable approximation ratio, also putting forward an online learning version under uncertain data volumes and network delays.
Even if we start from the same observation, our work proceeds in a different direction of estimating the energy consumption, which is very important in \acp{EDN}, under the assumption of a simpler orchestration policy that allows us to focus more specifically on the comparison between stateless and stateful frameworks.
On the other hand, in~\cite{younis_energy-latency_2024} the authors put under the spotlight the trade-off between energy consumption and latency, which is addressed through a distributed algorithm that optimizes at the same time the task offloading decisions, local CPU frequency, and approximate computing accuracy, and is proven to be effective in a small-scale testbed and with numerical experiments.
A similar problem has been studied in~\cite{aghapour_task_2023}, though the decision-making algorithm uses deep reinforcement learning.
However, these papers assume a microservice, not serverless, architecture: this makes the contribution orthogonal to ours, albeit possibly relevant for further investigation.
Serverless is instead the specific subject of~\cite{russo_serverledge_2023}, where the authors design a FaaS platform for the edge-cloud continuum, called \emph{Serverledge}, supporting vertical/horizontal computation offloading.
The solution proposed is a possible implementation of the stateless deployment option discussed in this paper, without state management.

\section{Conclusions}\label{sec:conclusions}

In this paper, we have studied two alternative deployment options to realize stateful FaaS in \acp{EDN}.
The first relies on stateful FaaS runners interconnected via a service mesh of brokers to dispatch function invocations, where the applications' state resides on an external service accessed as on demand.
The second one exploits WebAssembly to assign one runner to each application instance, state included, thus directly realizing stateful FaaS.
Results from a testbed with small devices showed no noticeable drop in performance with several runners.
We have then defined models to estimate the energy consumption of processing and network traffic for the two deployment options, with known statistical characterization of the applications (arrival process, \ac{DAG} task dependencies, and state/invocation sizes).
For stateful FaaS, we have proposed a simple, yet effective, heuristic for allocating tasks to nodes upon application arrival and periodically to reduce fragmentation of edge node resources.
Extensive simulations have shown that stateful FaaS is more efficient than stateless FaaS, except for minimal applications' state or negligible network energy consumption compared to processing.

\optional{This work can be extended in many directions. First, it is possible to enhance the accuracy of the system model proposed by relaxing any combination of its assumptions, including the following: the network traffic energy does not depend on the source/destination, which may be incorrect for complex/irregular network topologies, especially in a decentralized environment; there are unlimited network resources, in terms of bandwidth, which can be optimistic when allocating data-intensive applications on isolated edge nodes; each task can be assigned to any node, without considering limitations such as memory requirements or affinities with specific hardware capabilities, e.g., availability of GPUs.
Furthermore, more sophisticated simulations could be run to measure other \acp{KPI} of interest for the infrastructure operator (e.g., the network throughput and load of nodes), the service provider (e.g., fairness across applications), or the user (e.g., application latency), in addition to energy consumption, which is the main focus of this paper. 
Finally, large-scale testbed evaluation, with realistic applications and direct energy measurement, would be needed to validate the simulation results and conclusions drawn.}

\section*{Acknowledgment}
The work of A.~Passarella was funded by the European Union under the Italian National Recovery and Resilience Plan (NRRP) of NextGenerationEU, partnership on “Telecommunications of the Future” (PE00000001 -- program ``RESTART''). The work was partly funded by the European Union under the projects EDGELESS (GA no. 101092950), 6Green (GA 101096925), and GreenDIGIT (GA 101131207).

\begin{acronym}
  \acro{3GPP}{Third Generation Partnership Project}
  \acro{5G-PPP}{5G Public Private Partnership}
  \acro{5GC}{5G Core}
  \acro{AA}{Authentication and Authorization}
  \acro{ADF}{Azure Durable Function}
  \acro{AI}{Artificial Intelligence}
  \acro{API}{Application Programming Interface}
  \acro{AP}{Access Point}
  \acro{AR}{Augmented Reality}
  \acro{BGP}{Border Gateway Protocol}
  \acro{BSP}{Bulk Synchronous Parallel}
  \acro{BS}{Base Station}
  \acro{CDF}{Cumulative Distribution Function}
  \acro{CFS}{Customer Facing Service}
  \acro{CPU}{Central Processing Unit}
  \acro{DAG}{Directed Acyclic Graph}
  \acro{DHT}{Distributed Hash Table}
  \acro{DNS}{Domain Name System}
  \acro{EAS}{Edge Application Server}
  \acro{ECSP}{Edge Computing Service Provider}
  \acro{EDN}{Edge Data Network}
  \acro{ETSI}{European Telecommunications Standards Institute}
  \acro{FCFS}{First Come First Serve}
  \acro{FSM}{Finite State Machine}
  \acro{FaaS}{Function as a Service}
  \acro{GPU}{Graphics Processing Unit}
  \acro{HTML}{HyperText Markup Language}
  \acro{HTTP}{Hyper-Text Transfer Protocol}
  \acro{ICN}{Information-Centric Networking}
  \acro{ICT}{Information and Communication Technologies}
  \acro{IETF}{Internet Engineering Task Force}
  \acro{IIoT}{Industrial Internet of Things}
  \acro{ILP}{Integer Linear Programming}
  \acro{IPP}{Interrupted Poisson Process}
  \acro{IP}{Internet Protocol}
  \acro{ISG}{Industry Specification Group}
  \acro{ITS}{Intelligent Transportation System}
  \acro{ITU}{International Telecommunication Union}
  \acro{IT}{Information Technology}
  \acro{IaaS}{Infrastructure as a Service}
  \acro{IoT}{Internet of Things}
  \acro{JSON}{JavaScript Object Notation}
  \acro{K8s}{Kubernetes}
  \acro{KPI}{Key Performance Indicator}
  \acro{KVS}{Key-Value Store}
  \acro{LCM}{Life Cycle Management}
  \acro{LL}{Link Layer}
  \acro{LTE}{Long Term Evolution}
  \acro{MAC}{Medium Access Layer}
  \acro{MBWA}{Mobile Broadband Wireless Access}
  \acro{MCC}{Mobile Cloud Computing}
  \acro{MEC}{Multi-access Edge Computing}
  \acro{MEH}{Mobile Edge Host}
  \acro{MEPM}{Mobile Edge Platform Manager}
  \acro{MEP}{Mobile Edge Platform}
  \acro{ME}{Mobile Edge}
  \acro{ML}{Machine Learning}
  \acro{MNO}{Mobile Network Operator}
  \acro{NAT}{Network Address Translation}
  \acro{NFV}{Network Function Virtualization}
  \acro{NFaaS}{Named Function as a Service}
  \acro{NN}{Neural Network}
  \acro{OSPF}{Open Shortest Path First}
  \acro{OSS}{Operations Support System}
  \acro{OS}{Operating System}
  \acro{OWC}{OpenWhisk Controller}
  \acro{PMF}{Probability Mass Function}
  \acro{PU}{Processing Unit}
  \acro{PaaS}{Platform as a Service}
  \acro{PoA}{Point of Attachment}
  \acro{QoE}{Quality of Experience}
  \acro{QoS}{Quality of Service}
  \acro{RAN}{Radio Access Network}
  \acro{RPC}{Remote Procedure Call}
  \acro{RR}{Round Robin}
  \acro{RSU}{Road Side Unit}
  \acro{SBC}{Single-Board Computer}
  \acro{SDN}{Software Defined Networking}
  \acro{SJF}{Shortest Job First}
  \acro{SLA}{Service Level Agreement}
  \acro{SMP}{Symmetric Multiprocessing}
  \acro{SoC}{System on Chip}
  \acro{SRPT}{Shortest Remaining Processing Time}
  \acro{SPT}{Shortest Processing Time}
  \acro{STL}{Standard Template Library}
  \acro{SaaS}{Software as a Service}
  \acro{TCP}{Transmission Control Protocol}
  \acro{TPU}{Tensor Processing Unit}
  \acro{TSN}{Time-Sensitive Networking}
  \acro{UDP}{User Datagram Protocol}
  \acro{UE}{User Equipment}
  \acro{UPF}{User Plane Function}
  \acro{URI}{Uniform Resource Identifier}
  \acro{URL}{Uniform Resource Locator}
  \acro{UT}{User Terminal}
  \acro{VANET}{Vehicular Ad-hoc Network}
  \acro{VIM}{Virtual Infrastructure Manager}
  \acro{VR}{Virtual Reality}
  \acro{VM}{Virtual Machine}
  \acro{VNF}{Virtual Network Function}
  \acro{WLAN}{Wireless Local Area Network}
  \acro{WMN}{Wireless Mesh Network}
  \acro{WRR}{Weighted Round Robin}
  \acro{YAML}{YAML Ain't Markup Language}
\end{acronym}





\end{document}